\documentclass[hyper,a4paper,12pt,oneside]{JHEP}

\usepackage{amsmath}
\usepackage{graphicx}
\usepackage[dvips]{psfrag}

\newcommand{\bz}{\bar{z}}
\newcommand{\cd}{c^{\dag}}
\newcommand{\bra}[2]{\left<#1,#2\right|}
\newcommand{\cket}[2]{\left|#1,#2\right>}
\newcommand{\nh}{\hat{n}}
\newcommand{\Nh}{\hat{N}}
\newcommand{\Db}{\bar{D}}
\newcommand{\alb}{\bar{\alpha}}
\newcommand{\Deltah}{\hat{\Delta}}
\newcommand{\uh}{\hat{u}}
\newcommand{\wh}{\hat{w}}

\title{Elongated U(1) Instantons on Noncommutative $\bf R^4$}

\author{Tomomi Ishikawa, Shin-Ichiro Kuroki and Akifumi Sako\\
        Graduate School of Science, Hiroshima University,\\
        1-3-1 Kagamiyama, Higashi-Hiroshima 739-8526, Japan\\
        E-mail: \email{tomomi@theo.phys.sci.hiroshima-u.ac.jp},\\
        \hspace{14mm}\email{kuroki@theo.phys.sci.hiroshima-u.ac.jp},\\
        \hspace{14mm}\email{sako@math.sci.hiroshima-u.ac.jp}\\}

\preprint{HUPD-0113\\ \hepth{0109111}}

\abstract{We found an exact solution of elongated U(1) instanton
	  on noncommutative $\bf{R}^4$ for general instanton number $k$.
	  The deformed ADHM equation was solved with general $k$ and
	  the gauge connection and the curvature were given explicitly.
	  We also checked our solutions and evaluated the instanton
	  charge by a numerical calculation.}
	  

\begin{document}

\section{Introduction}

Recently, noncommutative geometry and noncommutative field theory
have been revived in the string theory and are studied actively
\cite{Connes}\cite{Seiberg}\cite{Douglas}.
In this concern, noncommutative instantons have been discussed
for recent three years.
Especially U(1) gauge group case is analyzed in detail and
there are some concrete form of gauge field whose curvature
is (anti-)self-dual.
However, these explicit form is given only at the case of instanton
number $k=1$ and $2$ \cite{Nekrasov3}\cite{Nekrasov4}\cite{Kim}\cite{Chu}.
The gauge field of a general $k$ instanton had never been constructed
for the reason of the difficulty of the calculation without
some special cases.
For example, when the noncommutative parameter is given as anti-self-dual,
then the solutions of the anti-self-dual equation of the curvature are
obtained by simple solution generating technique
\cite{Harvey}\cite{Aganatic}\cite{Hamanaka}.
These solutions (called localized instanton) correspond to the case that
the stability condition of moduli space is removed \cite{Furuuchi3}.
Therefore we have to construct an explicit expression of the instanton
satisfying stability condition\footnote{In the Braden and Nekrasov's
paper \cite{Nekrasov2}, they give
the elongated instanton solution for general $k$ as the solution of
deformed ADHM equations. The solution contains the Charlier polynomials,
and corresponds to commutative space instanton with nontrivial metric.
On the other hand, our elongated instantons are given in noncommutative
$\bf R^4$ (\S\ref{SEC:elongated}).
Therefore our solutions are different from their one.}.
If an exact expression of a such gauge field and its curvature are given,
they will be powerful tool for many kind of calculations.
For example, in cohomological Yang-Mills theory we had to evaluate sum of
Euler number of instanton moduli space for each instanton number.
In the similar case of noncommutative Yang-Mills,
the explicit expression of gauge field is expected to be a more powerful
tool since we have similar example in GMS solitons
\cite{GMS}\cite{Sako1}\cite{Sako2}.\\

In this paper, noncommutative U(1) instantons with arbitrary instanton
number $k$ are constructed with deformed ADHM procedure.
The gauge connections and curvature 2-form are given as a concrete form
with number operator (Fock space) representation at the same time.\\

Another motivation of this paper is to resolve the issue that instanton
number in noncommutative ADHM construction do not correspond to
the integral of the first Pontrjagin class (instanton charge) with clear way,
where we call the size of matrices in ADHM construction instanton number.
Recently, some new results about the instanton number of the
noncommutative $k=2$ U(1) instanton is given \cite{Kim}\cite{Chu}.
We also analyze higher instanton number case in \S\ref{SEC:numerical}
by a numerical way.

\section{Noncommutative U(1) Instantons}

\subsection{Noncommutative $\bf{R}^4$ and the Fock space representation}

We consider Euclidean noncommutative $\bf{R}^4$, whose coordinates
$x^{\mu}\;(\mu=1,2,3,4)$ satisfy the commutation relations
\begin{equation}
 [x^{\mu},x^{\nu}]=i\theta^{\mu\nu}, \label{EQ:xx-commu}
\end{equation}
where $\theta^{\mu\nu}$ is an antisymmetric real constant matrix,
and is called noncommutative parameter.
We can always bring $\theta^{\mu\nu}$ to the skew-diagonal form
\begin{equation}
 \theta^{\mu\nu}=
  \begin{pmatrix}
   0            & \theta^{12} & 0            & 0 \\
   -\theta^{12} & 0           & 0            & 0 \\
   0            & 0           & 0            & \theta^{34} \\
   0            & 0           & -\theta^{34} & 0
  \end{pmatrix}
\end{equation}
by space rotation.
For simplicity, we restrict the noncommutativity of space to the
case of $\theta^{12}=\theta^{34}=-\zeta\;(\zeta>0)$.
Here we introduce complex coordinates
\begin{equation}
 z_1=\frac{1}{\sqrt{2}}(x^1+ix^2),\;\; z_2=\frac{1}{\sqrt{2}}(x^3+ix^4),
  \label{EQ:z-coordinates}
\end{equation}
then the commutation relations (\ref{EQ:xx-commu}) become
\begin{equation}
 [z_1,\bz_1]=[z_2,\bz_2]=-\zeta,\;\;\mbox{others are zero}.
  \label{EQ:z-commu}
\end{equation}
For using the usual operator representation, we define creation and
annihilation operators by
\begin{equation}
 \cd_{\alpha}=\frac{z_{\alpha}}{\sqrt{\zeta}},\;\;
  c_{\alpha}=\frac{\bz_{\alpha}}{\sqrt{\zeta}},\;\;
  [c_{\alpha},\cd_{\alpha}]=1\;\;\;\;(\alpha=1,2).\label{EQ:c-a-operator}
\end{equation}
The Fock space $\cal H$ on which the creation and annihilation
operators (\ref{EQ:c-a-operator}) act is spanned by the Fock states
\begin{equation}
 \left|n_1,n_2\right>
  =\frac{(\cd_1)^{n_1}(\cd_2)^{n_2}}{\sqrt{n_1!n_2!}}\left|0,0\right>,
\end{equation}
with
\begin{eqnarray}
 c_1\cket{n_1}{n_2}=\sqrt{n_1}\cket{n_1-1}{n_2},\;\;&&
  \cd_1\cket{n_1}{n_2}=\sqrt{n_1+1}\cket{n_1+1}{n_2},\nonumber\\
 c_2\cket{n_1}{n_2}=\sqrt{n_2}\cket{n_1}{n_2-1},\;\;&&
  \cd_2\cket{n_1}{n_2}=\sqrt{n_2+1}\cket{n_1}{n_2+1},
\end{eqnarray}
where $n_1$ and $n_2$ are the occupation number.
The number operators are also defined by
\begin{equation}
 \nh_{\alpha}=\cd_{\alpha}c_{\alpha},\;\;\Nh=\nh_1+\nh_2,
\end{equation}
which act on the Fock states as
\begin{equation}
 \nh_{\alpha}\cket{n_1}{n_2}=n_{\alpha}\cket{n_1}{n_2},\;\;
  \Nh\cket{n_1}{n_2}=(n_1+n_2)\cket{n_1}{n_2}.
\end{equation}
In the operator representation, derivatives of a function
$f(z_1,\bz_1,z_2,\bz_2)$ are defined by
\begin{equation}
 \partial_{\alpha}f(z)=\frac{1}{\zeta}[\bz_{\alpha},f(z)],\;\;
  \bar{\partial}_{\alpha}f(z)=-\frac{1}{\zeta}[z_{\alpha},f(z)].
\end{equation}
The integral on noncommutative $\bf{R}^4$ is defined by the standard
trace in the operator representation,
\begin{equation}
 \int d^4x=\int d^4z=(2\pi\zeta)^2\mbox{Tr}.
\end{equation}
Note that $\mbox{Tr}$ represents the trace over the Fock
space whereas the trace over the gauge group is represented by
$\mbox{tr}$.

\subsection{Noncommutative gauge theory and instantons}

Here we consider a U(N) Yang-Mills theory on noncommutative $\bf R^4$.

In the noncommutative space, the Yang-Mills connection is defined as
\cite{Nekrasov3}\cite{Nekrasov4}\cite{Gross1}
\begin{equation}
 \nabla_{\mu}\Psi=i\Psi\theta_{\mu\nu}x^{\nu}+D_{\mu}\Psi,
\end{equation}
where $\Psi$ is matter field and $D_{\mu}$ is the gauge field
which is defined as anti-hermitian.
The Yang-Mills curvature of the connection $\nabla_{\mu}$ is
\begin{equation}
 F_{\mu\nu}=[\nabla_{\mu},\nabla_{\nu}]
  =-i\theta_{\mu\nu}+[D_{\mu},D_{\nu}].\label{EQ:F-x}
\end{equation}
In our notation of the complex coordinates (\ref{EQ:z-coordinates}) and
(\ref{EQ:z-commu}), the curvatures (\ref{EQ:F-x}) are
\begin{equation}
 F_{\alpha\alb}=\frac{1}{\zeta}+[D_{\alpha},\Db_{\alb}],\;\;
  F_{\alpha\bar{\beta}}=[D_{\alpha},\Db_{\bar{\beta}}]\;\;\;\;
  (\alpha\not=\beta).
\end{equation}
The Yang-Mills action is given by
\begin{equation}
 S=-\frac{1}{g^2}\int\mbox{tr}_NF\wedge *F,\label{EQ:Y-M-action}
\end{equation}
where $\mbox{tr}_N$ represents a trace for the gauge group U(N),
$g$ is the Yang-Mills coupling and $*$ is Hodge-star.
Its equation of motion is
\begin{equation}
 [\nabla_{\mu},F_{\mu\nu}]=0.\label{EQ:Y-M-eom}
\end{equation}
(Anti-)instanton solutions are special solutions of (\ref{EQ:Y-M-eom})
which satisfy the (anti)-self-duality condition
\begin{equation}
 F=\pm*F.
\end{equation}
(Anti-)self-duality conditions in the complex coordinates are
\begin{eqnarray}
 F_{1\bar{1}}=&+&F_{2\bar{2}},\;\;F_{1\bar{2}}=F_{\bar{1}2}=0\;\;\;\;
  \mbox{(self-dual)},\label{EQ:SD}\\
 F_{1\bar{1}}=&-&F_{2\bar{2}},\;\;F_{12}=F_{\bar{1}\bar{2}}=0\;\;\;\;
  \mbox{(anti-self-dual)}.\label{EQ:ASD}
\end{eqnarray}
In the commutative space, these solutions are classified by
the topological charge
\begin{equation}
 Q=-\frac{1}{8\pi^2}\int\mbox{tr}_NF\wedge F,\label{EQ:Q}
\end{equation}
which is always integer and is called instanton number $k$.
However, in the noncommutative space this statement is unclear.
We discuss this issue in \S\ref{SEC:numerical} by using the operator
representation of (\ref{EQ:Q}):
\begin{equation}
 Q=\begin{cases}
    -\zeta^2\mbox{Tr}\,\mbox{tr}_N
    (F_{1\bar{1}}F_{\bar{1}1}+F_{12}F_{\bar{1}\bar{2}})\;\;\;\;&
    \mbox{(self-dual)}\\
    \zeta^2\mbox{Tr}\,\mbox{tr}_N
    (F_{1\bar{1}}F_{\bar{1}1}+F_{1\bar{2}}F_{\bar{1}2})\;\;\;\;&
    \mbox{(anti-self-dual)}.
    \end{cases}\label{EQ:Q-op-N}
\end{equation}

\subsection{Nekrasov-Schwarz noncommutative U(1) instantons}

In the ordinary commutative space, there is a well-known way to find
(anti)-self-dual configurations of the gauge fields.
It is the ADHM construction which is proposed by Atiyah, Drinfeld, Hitchin
and Manin \cite{ADHM}.
Nekrasov and Schwarz first extended this method to noncommutative
space \cite{Nekrasov1}.
Especially U(1) case is discussed in
\cite{Nekrasov2}\cite{Nekrasov3}\cite{Nekrasov4}
\cite{Furuuchi1}\cite{Furuuchi2} in detail.\\

In commutative space case, the U(1) instanton is impossible to exist.
However, in noncommutative space case, nontrivial U(1) instantons exist.
Here we show a brief review on the ADHM construction of U(1) instanton
\cite{Nekrasov3}\cite{Nekrasov4}.

The first step of the ADHM construction is looking for matrices
$B_1$, $B_2$, $I$ and $J$ which satisfy the deformed ADHM equations
\begin{eqnarray}
 &&[B_1,B_1^{\dag}]+[B_2,B_2^{\dag}]+II^{\dag}-J^{\dag}J=2\zeta,
  \label{EQ:ADHM1}\\
 &&[B_1,B_2]+IJ=0,\label{EQ:ADHM2}
\end{eqnarray}
where $B_1$ and $B_2$ are $k\times k$ complex matrices,
$I$ and $J^{\dag}$ are $k\times 1$ complex matrices.
In U(1) case, if $\zeta >0$, Eq.(\ref{EQ:ADHM2}) and the stability
condition allows us to take $J=0$ \cite{Nakajima}\cite{Furuuchi2}.
In \cite{Nekrasov3}\cite{Nekrasov4}, a projector which projects the
Hilbert space $\cal H$ to the subspace of $\cal H$ is given as
\begin{equation}
 P=I^{\dag}e^{\sum_{\alpha}\beta_{\alpha}^{\dag}\cd_{\alpha}}
  \cket{0}{0}G^{-1}\bra{0}{0}
  e^{\sum_{\alpha}\beta_{\alpha}c_{\alpha}}I,\label{EQ:projector-N}
\end{equation}
where $B_{\alpha}=\sqrt{\zeta}\beta_{\alpha}$ and
$G$ is a normalization factor (hermitian matrix)
\begin{equation}
 G=\bra{0}{0}e^{\sum_{\alpha}\beta_{\alpha}c_{\alpha}}II^{\dag}
  e^{\sum_{\alpha}\beta_{\alpha}^{\dag}\cd_{\alpha}}\cket{0}{0}.
\end{equation}
We introduce a shift operator $S$ which satisfies
\begin{equation}
 SS^{\dag}=1,\;\;S^{\dag}S=1-P.
\end{equation}
Using the shift operator $S$, the U(1) anti-self-dual gauge fields
are given as
\begin{equation}
 D_{\alpha}=\sqrt{\frac{1}{\zeta}}S\Lambda^{-\frac{1}{2}}c_{\alpha}
  \Lambda^{\frac{1}{2}}S^{\dag},\;\;
  \Db_{\alb}=-\sqrt{\frac{1}{\zeta}}S\Lambda^{\frac{1}{2}}\cd_{\alpha}
  \Lambda^{-\frac{1}{2}}S^{\dag},\label{EQ:gauge-NS}
\end{equation}
where
\begin{equation}
 \Lambda=1+I^{\dag}\Deltah^{-1}I,\;\;
  \Deltah=\zeta\sum_{\alpha}(\beta_{\alpha}-\cd_{\alpha})
  (\beta_{\alpha}^{\dag}-c_{\alpha}).\label{EQ:Lambda}
\end{equation}
In the following section, we construct an explicit form of the gauge
fields (\ref{EQ:gauge-NS}).

\section{Construction of elongated U(1) instantons \label{SEC:elongated}}

We know the ADHM construction, but in usual,
it is hardly difficult to write down explicitly the solutions
of instantons for arbitrary instanton number $k$ even in the U(1) case.
However, in the case of the elongated U(1) instantons,
we can get the explicit expression for general $k$ instanton.\\

Let matrices $B_1$, $B_2$, $I$ and $J$ be
\begin{equation}
 B_1=\sum_{i=1}^{k-1}\sqrt{2i\zeta}e_ie_{i+1}^{\dag},\;\;
  B_2=0,\;\;I=\sqrt{2k\zeta}e_k,\;\;J=0,\label{EQ:B1B2IJ}
\end{equation}
where $e_i$ is defined as
\begin{equation}
 e_i^{\dag}=
 (\stackrel{1}{0},\stackrel{}{\cdots},\stackrel{i-1}{0},\stackrel{i}{1},
 \stackrel{i+1}{0},\stackrel{}{\cdots},\stackrel{k}{0}).
\end{equation}
These matrices (\ref{EQ:B1B2IJ}) satisfy the deformed ADHM equations
(\ref{EQ:ADHM1}) and (\ref{EQ:ADHM2}).
These ADHM data correspond to the configuration
that $k$ instantons are elongated into $z_1$-$\bz_1$ direction.
Next step to get the explicit expression of the gauge field
(\ref{EQ:gauge-NS}) is to find the explicit form of $\Lambda$
in (\ref{EQ:Lambda}).
In our case, $\Deltah$ in (\ref{EQ:Lambda}) is
\begin{equation}
 \Deltah=\zeta\left[\Nh+\sum_{i=1}^{k-1}\left\{2ie_ie_i^{\dag}
     -\sqrt{2i}(c_1e_ie_{i+1}^{\dag}+c_1^{\dag}e_{i+1}e_i^{\dag})
     \right\}\right],
\end{equation}
and $\Lambda$ is
\begin{equation}
 \Lambda=1+2k\zeta\Deltah_{kk}^{-1},
\end{equation}
where $\Deltah_{kk}^{-1}$ is the $(k,k)$ component of the matrix
$\Deltah^{-1}$.
It is sufficient for getting $\Deltah_{kk}^{-1}$ that we know the $k$-th
line of $\Deltah^{-1}$ which is defined by $\Deltah^{-1}\Deltah=1$.
For this purpose, we define a vector $\bf\uh$ as
\begin{equation}
 {\bf\uh}^t=\frac{1}{\zeta}\sum_{i=1}^{k}\uh_ie_i^{\dag},
\end{equation}
and following calculation is performed:
\begin{eqnarray}
 {\bf\uh}^t\Deltah&=&\left\{\uh_1(\Nh+2)-\uh_2\cd_1\right\}e_1^{\dag}
  \nonumber\\
 & &+\sum_{i=2}^{k-1}\left\{-\sqrt{2(i-1)}\uh_{i-1}c_1
      +\uh_i(\Nh+2i)-\sqrt{2i}\uh_{i+1}\cd_1\right\}e_i^{\dag}
      \nonumber\\
 & &+\left\{-\sqrt{2(k-1)}\uh_{k-1}c_1+\uh_k\Nh\right\}e_k^{\dag}.
\end{eqnarray}
To satisfy the equation of $k$-th line of $\Deltah^{-1}\Deltah=1$,
the recurrence relation is imposed:
\begin{eqnarray}
 \uh_2\cd_1-\uh_1(\Nh+2)&=&0,\nonumber\\
 \sqrt{2i}\uh_{i+1}\cd_1-\uh_i(\Nh+2i)+\sqrt{2(i-1)}\uh_{i-1}c_1&=&0\;\;\;\;
  (2\leq i\leq k-1).\label{EQ:3-rec-u}
\end{eqnarray}
Then we obtain $\Deltah_{kk}^{-1}$ as
\begin{equation}
 \Deltah_{kk}^{-1}=\frac{1}{\zeta}\left\{-\sqrt{2(k-1)}\uh_{k-1}c_1
  +\uh_k\Nh\right\}^{-1}\uh_k.
\end{equation}
Now we substitute $\wh_i(\nh_1,\nh_2)$ for $\uh_i$:
\begin{equation}
 \uh_i=\wh_{i-1}(\nh_1,\nh_2)\frac{(\cd_1)^{k-i}}{\sqrt{2^{i-1}(i-1)!}},
\end{equation}
and rewrite the recurrence relation (\ref{EQ:3-rec-u}) as
\begin{eqnarray}
 w_1-(N-k+3)w_0&=&0,\nonumber\\
 w_{i+1}-(N+3i-k+3)w_i+2i(n_1+i-k+1)w_{i-1}&=&0\nonumber\\
 &&\hspace{-15mm}(1\leq i\leq k-2).\label{EQ:3-rec-w}
\end{eqnarray}
Note that $w_i(n_1,n_2)$ depends on only number operators, then its
inverse is formally given by $w_i^{-1}(n_1,n_2)$, where $n_1$ and $n_2$
are replaced by their eigen values.
We can fortunately solve the recurrence relation (\ref{EQ:3-rec-w}).
The generating function of $w_i$ is
\begin{equation}
 F(t)=(1-t)^{-n_1+n_2+k-1}(1-2t)^{-n_2-1}
  =\sum_{i=0}^{\infty}\frac{w_i}{i!}t^i.
\end{equation}
Then $w_i(n_1,n_2)$ is given as
\begin{equation}
 w_i(n_1,n_2)=\left.\left(\frac{d}{dt}\right)^iF(t)\right|_{t=0}.
\end{equation}
Using this $w_i(n_1,n_2)$, we can write $\Lambda(n_1,n_2)$ as
\begin{equation}
 \Lambda(n_1,n_2)=\frac{w_k(n_1,n_2)}{w_k(n_1,n_2)-2kw_{k-1}(n_1,n_2)}.
\end{equation}
Next step, we should determine a shift operator.
In our case the projector (\ref{EQ:projector-N}) is
\begin{equation}
 P=\sum_{n_1=0}^{k-1}\cket{n_1}{0}\bra{n_1}{0},
\end{equation}
then we can define the shift operator as
\begin{equation}
 S^{\dag}=\sum_{n_1=0}^{\infty}\cket{n_1+k}{0}\bra{n_1}{0}
  +\sum_{n_1=0}^{\infty}\sum_{n_2=1}^{\infty}\cket{n_1}{n_2}\bra{n_1}{n_2}.
\end{equation}
Using the above results, we can write down the elongated U(1) instanton
gauge fields on noncommutative $\bf R^4$ explicitly:
\begin{eqnarray}
 D_1=& &\sqrt{\frac{1}{\zeta}}\sum_{n_1=0}^{\infty}\sum_{n_2=0}^{\infty}
  \cket{n_1}{n_2}\bra{n_1+1}{n_2}d_1(n_1,n_2;k),\\
 \Db_{\bar{1}}=&-&
  \sqrt{\frac{1}{\zeta}}\sum_{n_1=0}^{\infty}\sum_{n_2=0}^{\infty}
  \cket{n_1+1}{n_2}\bra{n_1}{n_2}d_1(n_1,n_2;k),\\
 D_2=&&\sqrt{\frac{1}{\zeta}}\left\{
  \sum_{n_1=0}^{\infty}\cket{n_1}{0}\bra{n_1+k}{1}d_2(n_1,0;k)\right.
  \nonumber\\
 &&\hspace{1cm}+\left.\sum_{n_1=0}^{\infty}\sum_{n_2=1}^{\infty}
     \cket{n_1}{n_2}\bra{n_1}{n_2+1}d_2(n_1,n_2;k)\right\},\\
 \Db_{\bar{2}}=&-&\sqrt{\frac{1}{\zeta}}\left\{
  \sum_{n_1=0}^{\infty}\cket{n_1+k}{1}\bra{n_1}{0}d_2(n_1,0;k)\right.
 \nonumber\\
 &&\hspace{1cm}+\left.\sum_{n_1=0}^{\infty}\sum_{n_2=1}^{\infty}
     \cket{n_1}{n_2+1}\bra{n_1}{n_2}d_2(n_1,n_2;k)\right\},
 \label{EQ:D-elong}
\end{eqnarray}
where
\begin{eqnarray}
 d_1(n_1,n_2;k)&=&\begin{cases}
\sqrt{n_1+k+1}
\left\{\frac{\Lambda(n_1+k+1,0)}
{\Lambda(n_1+k,0)}\right\}^{\frac{1}{2}}
  &\mbox{$(n_2=0)$,}\\
\sqrt{n_1+1}
\left\{\frac{\Lambda(n_1+1,n_2)}
{\Lambda(n_1,n_2)}\right\}^{\frac{1}{2}}
&\mbox{$(n_2\not=0)$,}
                \end{cases}\label{EQ:d1}\\
 d_2(n_1,n_2;k)&=&\begin{cases}
\left\{\frac{\Lambda(n_1+k,1)}
{\Lambda(n_1+k,0)}\right\}^{\frac{1}{2}}\hspace{25mm}
&\mbox{$(n_2=0)$,}\\
\sqrt{n_2+1}
\left\{\frac{\Lambda(n_1,n_2+1)}
{\Lambda(n_1,n_2)}\right\}^{\frac{1}{2}}
&\mbox{$(n_2\not=0)$.}
                \end{cases}\label{EQ:d2}
\end{eqnarray}
Its curvature is presented in Appendix \ref{AP:curvature}.

\section{Numerical check and analysis \label{SEC:numerical}}

In \S\ref{SEC:elongated}, we have constructed the solution of the
elongated U(1) instantons with general instanton number $k$.
In this section we check and analyze our solution numerically.

\subsection{Anti-self-duality}

Since our solution would be an anti-self-dual configuration,
it should satisfy the condition (\ref{EQ:ASD}).
We can show that the condition $F_{12}=F_{\bar{1}\bar{2}}=0$
for every component $\cket{n_1}{n_2}\bra{n_1}{n_2}$,
and $F_{1\bar{1}}=-F_{2\bar{2}}$ on $\cket{0}{0}\bra{0}{0}$
component are satisfied with analytic calculation easily.
However, it is difficult to check the condition
$F_{1\bar{1}}=-F_{2\bar{2}}$ for any component analytically.
Therefore we check the condition numerically and
the results satisfy (\ref{EQ:ASD}) completely.

\subsection{Instanton number and density}

\begin{figure}[t]
  \begin{center}
           \scalebox{0.5}{\includegraphics{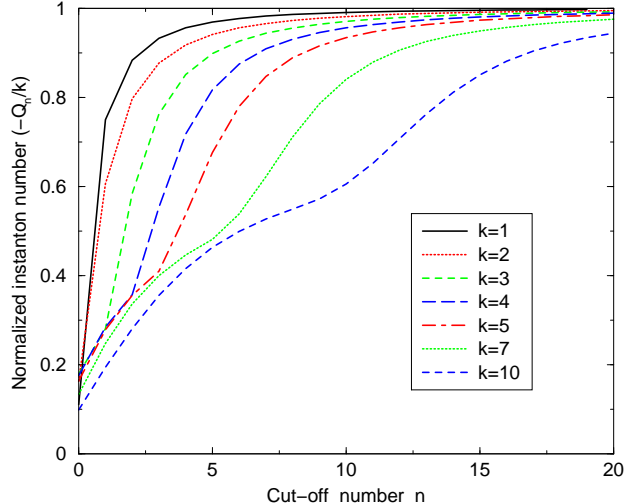}}
  \end{center}
\caption{Convergency of the instanton charge;
As the cut-off number $n$ increases, all the normalized
instanton number $-Q_n/k$ approaches one.
These results show that our solution is suitable and gives
the instanton number precisely.}
\label{Fig:conv}
\end{figure}

\begin{figure}[p]
 \begin{center}
  \begin{minipage}{70mm}
   \begin{center}
    \psfragscanon
    \psfrag{X1}[][][1.0]{$n_2$}
    \psfrag{X2}[][][1.0]{$n_1$}
    \psfrag{X3}[r][][1.0]{$\rho$}
    \psfrag{1}[][][0.8]{0}
    \psfrag{6}[][][0.8]{5}
    \psfrag{11}[][][0.8]{10}
    \psfrag{16}[][][0.8]{15}
    \psfrag{21}[][][0.8]{20}
    \scalebox{0.8}{\includegraphics{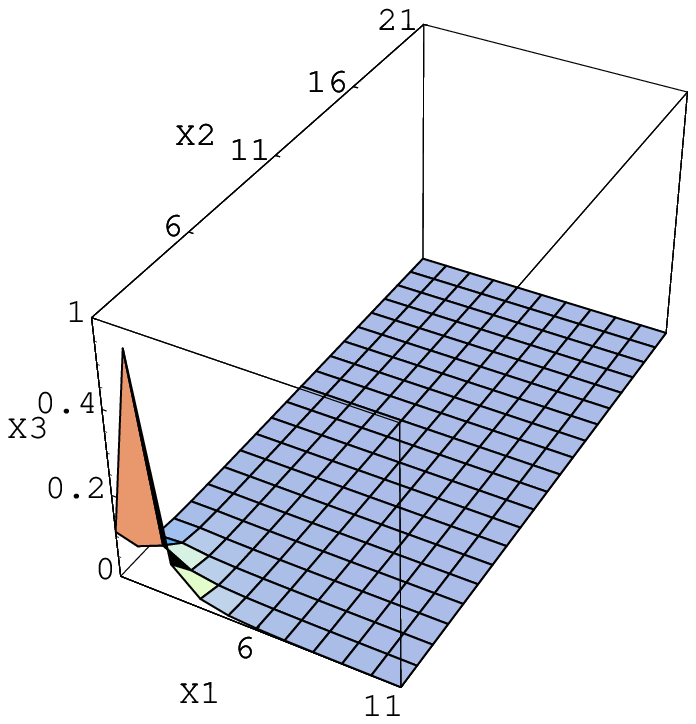}}
   \end{center}
   \caption{$k=1$ instanton-density; The height of the
   graph is the instanton density. The instanton density is localized
   around the maximum point $(n_1,n_2)=(0,1)$. At this maximum point the
   instanton density is $\rho=0.55556$.}
   \label{Fig:3D1}
  \end{minipage}
  \hspace{5mm}
  \begin{minipage}{70mm}
   \begin{center}
    \psfragscanon
    \psfrag{X1}[][][1.0]{$n_2$}
    \psfrag{X2}[][][1.0]{$n_1$}
    \psfrag{X3}[r][][1.0]{$\rho$}
    \psfrag{1}[][][0.8]{0}
    \psfrag{6}[][][0.8]{5}
    \psfrag{11}[][][0.8]{10}
    \psfrag{16}[][][0.8]{15}
    \psfrag{21}[][][0.8]{20}
    \scalebox{0.8}{\includegraphics{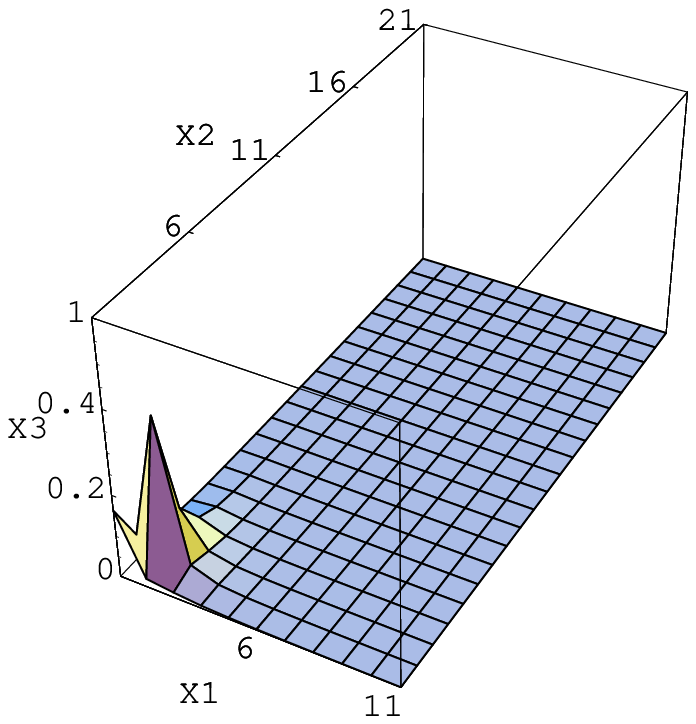}}
   \end{center}
   \caption{$k=2$ instanton density; At the point $(n_1,n_2)=(1,1)$
   the instanton density takes the maximum value $\rho=0.366429$.\vspace{11mm}}
   \label{Fig:3D2}
  \end{minipage}
 \end{center}
\end{figure}

\begin{figure}[p]
 \begin{center}
  \psfragscanon
  \psfrag{X1}[][][1.5]{$n_2$}
  \psfrag{X2}[][][1.5]{$n_1$}
  \psfrag{X3}[][][1.5]{$\rho$}
  \psfrag{1}[][][0.8]{0}
  \psfrag{6}[][][0.8]{5}
  \psfrag{31}[][][0.8]{30}
  \psfrag{61}[][][0.8]{60}
  \psfrag{0.05}[][][0.8]{}
 \psfrag{0.03}[][][0.8]{0.03}
  \psfrag{0.02}[][][0.8]{0.02}
  \psfrag{0.01}[][][0.8]{0.01}
   \psfrag{0}[][][0.8]{0}
  \scalebox{0.9}{\includegraphics{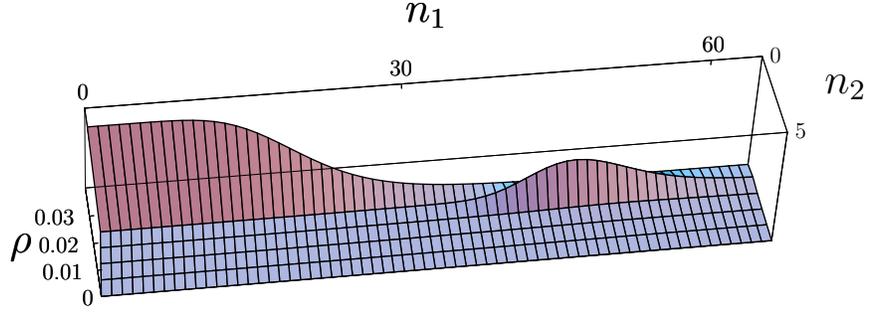}}
 \end{center}
 \caption{$k=30$ instanton density; At the origin the instanton density
 takes the maximum value $\rho=0.0333333$. And the another extreme value
 is $\rho=0.0127042$ at the point $(n_1,n_2)=(46,1)$.}
 \label{Fig:3D30}
\end{figure}

\begin{figure}[t]
  \begin{center}
   \psfragscanon
    \psfrag{X1}[][][1.5]{$n_2$}
    \psfrag{X2}[][][1.5]{$n_1$}
      \psfrag{X3}[][][1.5]{$\rho$}
    \psfrag{1}[][][0.8]{0}
     \psfrag{6}[][][0.8]{5}
    \psfrag{11}[][][0.8]{30}
     \psfrag{21}[][][0.8]{60}
      \psfrag{31}[][][0.8]{90}
   \psfrag{41}[][][0.8]{120}
    \psfrag{51}[][][0.8]{150}
     \psfrag{61}[][][0.8]{180}
    \psfrag{0.01}[][][0.8]{0.01}  
         \scalebox{0.9}{\includegraphics{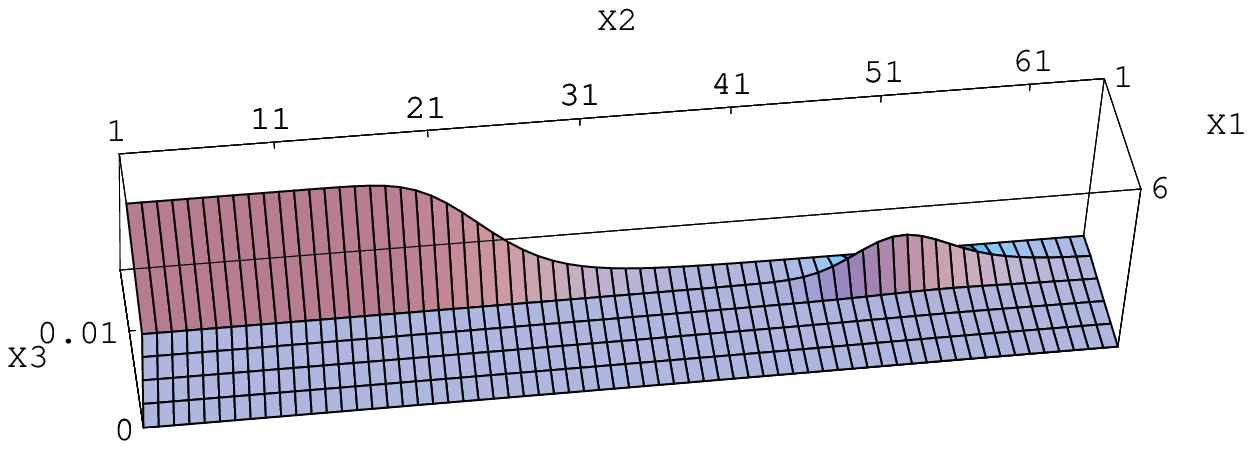}}
  \end{center}
\caption{$k=90$ instanton density: On the $n_2=0$ line the instanton density
 near the origin takes the maximum value $\rho=0.0111111$.
 And the another extreme value is $\rho=0.00397892$
 at the point $(n_1,n_2)=(157, 1)$.} 
\label{Fig:3D90}
\end{figure}

\begin{table}[t]
 \begin{center}
 \begin{minipage}{85mm}
  \begin{center}
  \begin{tabular}{c|ccc}\hline
   &               & $Q_n$  &\\
   \;k\; &$n=10$  & $n=20$      & $n=50     $\\ \hline
 1  & $-0.991033$ & $-0.997442$ & $-0.999557$ \\
 2  & $-1.96443$  & $-1.98977$  & $-1.99823$  \\
 3  & $-2.91641$  & $-2.97653$  & $-2.99599$  \\
 4  & $-3.83934$  & $-3.95662$  & $-3.99282$  \\
 5  & $-4.72416$  & $-4.92835$  & $-4.98864$  \\
 6  & $-5.56423$  & $-5.88927$  & $-5.98336$  \\
 7  & $-6.35653$  & $-6.83624$  & $-6.97688$  \\
 8  & $-7.09772$  & $-7.76553$  & $-7.96905$  \\
 9  & $-7.79167$  & $-8.67337$  & $-8.95968$  \\
 10 & $-8.45736$  & $-9.55689$  & $-9.94857$  \\
 15 & $-11.1219$  & $-13.6201$  & $-14.855$   \\ \hline
  \end{tabular}
 \caption{Numerical check of $Q$ (i):
 $k$ is the instanton number.
   We summed over the Fock space up to $(n_1,n_2)=(10,10),(20,20),(50,50)$.}
 \label{TAB:Q-1}
\end{center}
\end{minipage}
\hspace{5mm}
 \begin{minipage}{55mm}
  \begin{center}
 \begin{tabular}{c|c}\hline
  \;$k$\;& $n$ of $99\%$ line\\ \hline
  1  & $10$ \\
  2  & $14$ \\
  3  & $18$ \\
  4  & $21$ \\
  5  & $24$ \\
  6  & $27$ \\
  7  & $30$ \\
  8  & $33$ \\
  9  & $35$ \\
  10 & $38$ \\
  15 & $50$ \\ \hline
 \end{tabular}
 \caption{Numerical check of $Q$ (ii):
   $n$ of $99\%$ line denotes the cut-off number $n$ at which $-Q_n/k=0.9$.}
   \label{TAB:Q-2}
   \end{center}
 \end{minipage}
\end{center}
\end{table}
In this subsection, we give the numerical study on our instanton solution.
The instanton charge (\ref{EQ:Q-op-N}) is given by
the infinite series; the trace operation is performed by the sum over
the Fock states $|n_1, n_2\rangle$, or the space-time points
$(n_1,n_2)$.
Since the sum contains the infinite points, in the numerical analysis
we restrict the number of space-time points, that is, we put
the cut-off number for the Infrared scale.
Then we define the instanton charge with the cut-off number $n$;
\begin{eqnarray}
 Q_n&=&-k\sum^n_{n_1=0}\sum^n_{n_2=0}\rho(n_1,n_2),\\
 Q_{\infty}&=&\lim_{n\to\infty}Q_n=Q,
\end{eqnarray}
and  we define the instanton density as
\begin{eqnarray}
  \rho(n_1,n_2)
   &:=&
        \frac{\zeta ^2}{k}
         \langle n_1,n_2|
     (F_{1\bar{1}}F_{\bar{1}1}
      +F_{1\bar{2}}F_{\bar{1}2} )
               |n_1,n_2\rangle.
\end{eqnarray}
The numerical results of the instanton charge is listed in the Table
\ref{TAB:Q-1}. At each instanton solution we give the three results;
the sum up to the cut-off number $n=10,20$ and $50$.
These results show that if we summed over the Fock space up to a higher
number $n$, the numerical calculation of the instanton charge approaches
the expected number. Fig.\ref{Fig:conv} shows the convergence of the
numerical calculation of the instanton charge.
In this figure we find that each normalized instanton number($-Q_n/k$)
approaches to one as we take a higher cut-off number $n$.
Then we conclude the convergency of the instanton charge of our solution
is suitable.
The Tab.\ref{TAB:Q-2} shows the cut-off number that gives $99\%$ of the
expected value. This $99\%$ cut-off number increases as the instanton
charge increases. This results mean that a high number instanton
solution is expanded into the space direction $(n_1, n_2)$. The
distributions of the instanton density are given in
Fig.\ref{Fig:3D1}$\sim$\ref{Fig:3D90}. The Fig.\ref{Fig:3D1} show
the instanton density of the one instanton solution. In this figure we
find that the configuration near the origin dominates (the maximum point
is $(n_1,n_2)=(0,1)$). As the instanton number increase the
configuration near the origin spreads out into the $n_1$ direction;
the point giving the extreme value moves along the $n_2=1$ line,
and the new maximum point appears from the origin
(Fig.\ref{Fig:3D30}$\sim$\ref{Fig:3D90}).
Our instanton solutions do not expand into the $n_2$ direction and
are localized around the $n_2$=0 plane ($|z_2|=0$ plane).
\begin{figure}[p]
 \begin{center}
  \begin{minipage}{70mm}
   \begin{center}
    \psfragscanon
    \psfrag{X1}[][][1.0]{$n_2$}
    \psfrag{X2}[][][1.0]{$n_1$}
    \psfrag{X3}[][][1.0]{$\rho$}
    \psfrag{1}[][][0.8]{0}
    \psfrag{6}[][][0.8]{5}
    \psfrag{11}[][][0.8]{10}
    \psfrag{16}[][][0.8]{15}
    \psfrag{21}[][][0.8]{20}
    \scalebox{0.8}{\includegraphics{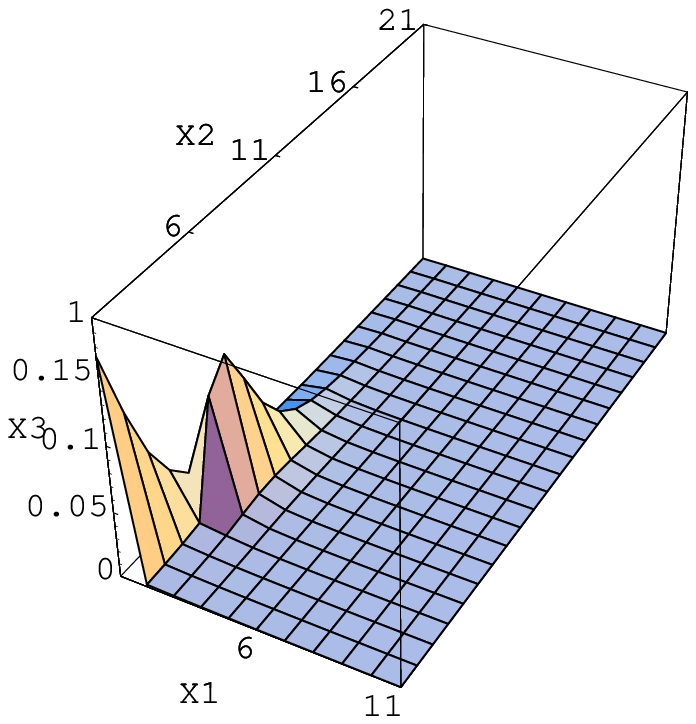}}
   \end{center}
   \caption{$k=5$ instanton density; At the origin the
   instanton density takes the maximum value $\rho=0.163719$.
   Another extreme value exists
   at the point $(n_1,n_2)=(5,1)$, and is $\rho=0.109048$.  }
   \label{Fig:3D5}
 \end{minipage}
 \hspace{5mm}
  \begin{minipage}{70mm}
   \begin{center}
    \psfragscanon
    \psfrag{X1}[][][1.0]{$n_2$}
    \psfrag{X2}[][][1.0]{$n_1$}
    \psfrag{X3}[][][1.0]{$\rho$}
   \psfrag{1}[][][0.8]{0}
    \psfrag{6}[][][0.8]{5}
    \psfrag{11}[][][0.8]{10}
    \psfrag{16}[][][0.8]{15}
    \psfrag{21}[][][0.8]{20}
   \psfrag{0}[][][0.8]{0}
   \scalebox{0.8}{\includegraphics{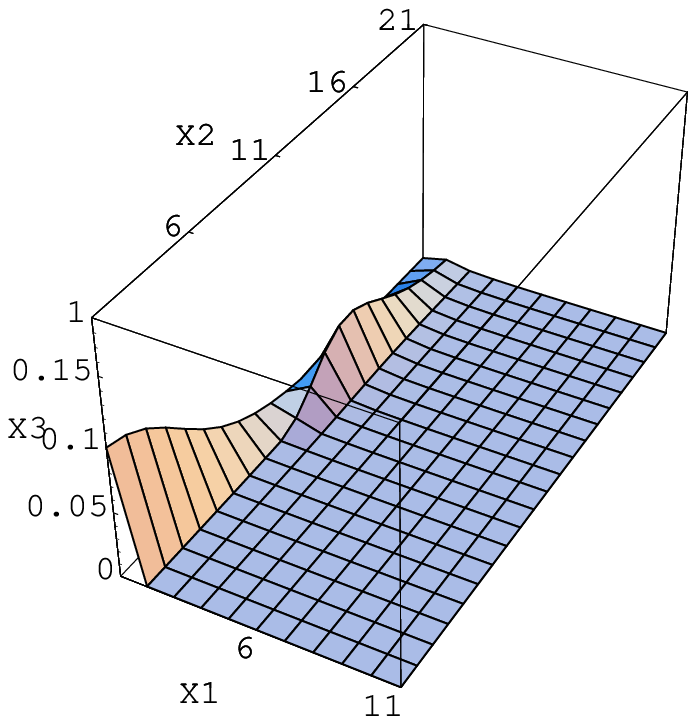}}
   \end{center}
   \caption{$k=10$ instanton density; At the origin the instanton
   density takes the maximum value $\rho=0.0989281$.
   Another extreme value is $\rho=0.0449079$
   at the point $(n_1,n_2)=(13,1)$.}
   \label{Fig:3D10}
  \end{minipage}
 \end{center}
\end{figure}
\begin{figure}[t]
  \begin{center}
   \psfragscanon
    \psfrag{X1}[][][1.5]{$n_2$}
    \psfrag{X2}[][][1.5]{$n_1$}
      \psfrag{X3}[][][1.5]{$\rho$}
    \psfrag{1}[][][0.8]{0}
     \psfrag{6}[][][0.8]{5}
   \psfrag{16}[][][0.8]{30}
    \psfrag{31}[][][0.8]{60}
   \psfrag{46}[][][0.8]{90}
     \psfrag{61}[][][0.8]{120}
      \psfrag{91}[][][0.8]{150}
       \psfrag{121}[][][0.8]{120}
     \psfrag{0.02}[][][0.8]{0.02}
      \psfrag{0.01}[][][0.8]{0.01}
       \psfrag{0}[][][0.8]{0}
                \scalebox{0.9}{\includegraphics{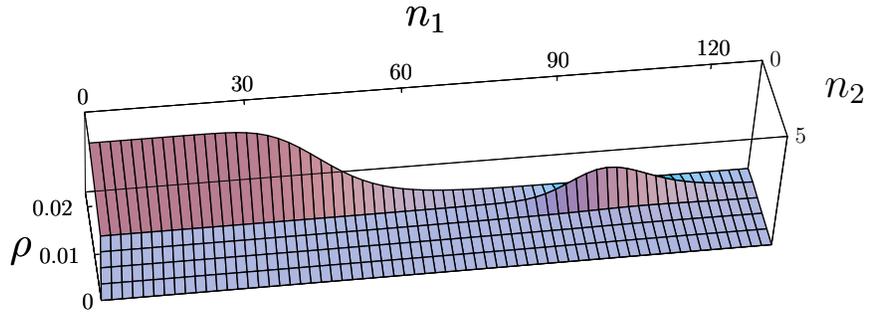}}
  \end{center}
\caption{$k=60$ instanton density:
 On the $n_2=0$ line, the instanton density near the origin 
 takes the maximum value $\rho=0.0166667$. And the another extreme value
 is $\rho=0.00610433$ at the point $(n_1,n_2)=(101,1)$.} 
\label{Fig:3D60}
\end{figure}

\section{Discussion and Conclusion}

Explicit expression of the elongated U(1) instanton
on noncommutative $\bf R^4$ was obtained for general instanton number.
Deformed ADHM construction was used there, and an important progress
is to construct exact form of the gauge field and the curvature.
As a result of this, we could perform two analysis of geometrical and
topological nature with numerical way.
First, the anti-self-dual condition is confirmed.
Second, numerical calculation of the instanton charge that defined by
Eq.(\ref{EQ:Q}) was done, and it implies that both instanton charge $Q$
defined by Eq.(\ref{EQ:Q}) and instanton number defined
by the size $k$ of the matrices of
ADHM construction is equivalent.
Additionally, we saw the distribution of instanton density of the
elongated instanton and it is indeed elongated to the $z_1$-$\bz_1$
direction.
From the point of view of Hilbert scheme, as Nakajima and Furuuchi said
in \cite{Nakajima}\cite{Furuuchi2},
$B_i$ determine the ideal.
In our case we take $B_2=0$ and $B_1\not= 0$, and off-diagonal part of
$B_1$ is regarded as the trace of k-points elongated to $z_1$-$\bz_1$
direction before shrinking into the origin.
This effect is still alive in density of the topological charge.\\

In these investigation, some geometrical natures of noncommutative
space appear.
For example, some kind of recurrence relations
plays a role of a differential equation on usual commutative spaces,
i.e., local geometry is written by some series.
Therefore the integration of some kind of topological charge
is replaced by the sum of series.
Unfortunately, many of topological charge are difficult to calculate
by analytic way.
This fact will demand to review the theory of sequence
from a quantum geometrical view point.

\acknowledgments

A.S is supported by JSPS Research Fellowships for Young Scientists.
Discussions during the YITP workshop YITP-W-01-04 ``Quantum Field Theory
2001'' were helpful to complete this work.

\appendix

\section{Curvature of our solution}\label{AP:curvature}

We present the curvature of the elongated U(1) instanton with general
instanton number $k$ by using the connection (\ref{EQ:D-elong}).
Our solution is  an anti-self-dual configuration, so
$F_{12}=F_{\bar{1}\bar{2}}=0$.
The other components are
\begin{eqnarray}
 F_{1\bar{1}}&=&\frac{1}{\zeta}-\frac{1}{\zeta}\left[
  \sum_{n_2=0}^{\infty}\cket{0}{n_2}\bra{0}{n_2}\bigl(d_1(0,n_2;k)\bigr)^2
  \right.
 \nonumber\\
 &&+\left.\sum_{n_1=1}^{\infty}\sum_{n_2=0}^{\infty}
  \cket{n_1}{n_2}\bra{n_1}{n_2}
  \Bigl\{\bigl(d_1(n_1,n_2;k)\bigr)^2-\bigl(d_1(n_1-1,n_2;k)\bigr)^2
  \Bigr\}\right],\nonumber\\
\end{eqnarray}
\begin{eqnarray}
 F_{2\bar{2}}&=&\frac{1}{\zeta}-\frac{1}{\zeta}\left[
  \sum_{n_1=0}^{\infty}\cket{n_1}{0}\bra{n_1}{0}
  \bigl(d_2(n_1,0;k)\bigr)^2\right.
 \nonumber\\
 &&+\sum_{n_1=0}^{k-1}\cket{n_1}{1}\bra{n_1}{1}
  \bigl(d_2(n_1,1;k)\bigr)^2\nonumber\\
 &&+\sum_{n_1=k}^{\infty}\cket{n_1}{1}\bra{n_1}{1}\Bigl\{
   \bigl(d_2(n_1,1;k)\bigr)^2-\bigl(d_2(n_1-k,0;k)\bigr)^2\Bigr\}\nonumber\\
 &&+\left.\sum_{n_1=0}^{\infty}\sum_{n_2=2}^{\infty}
     \cket{n_1}{n_2}\bra{n_1}{n_2}
  \Bigl\{\bigl(d_2(n_1,n_2;k)\bigr)^2-\bigl(d_2(n_1,n_2-1;k)\bigr)^2\Bigr\}
   \right],\nonumber\\ \\
 F_{1\bar{2}}&=&-\frac{1}{\zeta}\Biggl[
 \cket{k-1}{1}\bra{0}{0}d_1(k-1,1;k)d_2(0,0;k)\Biggr..\nonumber\\
 &&+\sum_{n_1=1}^{\infty}\cket{n_1+k-1}{1}\bra{n_1}{0}\nonumber\\
 &&\hspace{5mm}\times
  \Bigl\{d_1(n_1+k-1,1;k)d_2(n_1,0;k)-d_1(n_1-1,0;k)d_2(n_1-1,0;k)\Bigr\}
  \nonumber\\
 &&+\sum_{n_1=1}^{\infty}\sum_{n_2=1}^{\infty}
  \cket{n_1-1}{n_2+1}\bra{n_1}{n_2}\nonumber\\
 &&\hspace{5mm}\times
  \Bigl\{d_1(n_1-1,n_2+1;k)d_2(n_1,n_2;k)
  -d_1(n_1-1,n_2;k)d_2(n_1-1,n_2;k)\Bigr\}\Biggr],\nonumber\\ \\
 F_{\bar{1}2}&=&-F_{1\bar{2}}^{\dag}\nonumber\\
 &=&\frac{1}{\zeta}\Biggl[
 \cket{0}{0}\bra{k-1}{1}d_1(k-1,1;k)d_2(0,0;k)\Biggr.\nonumber\\
 &&+\sum_{n_1=1}^{\infty}\cket{n_1}{0}\bra{n_1+k-1}{1}\nonumber\\
 &&\hspace{5mm}\times
  \Bigl\{d_1(n_1+k-1,1;k)d_2(n_1,0;k)-d_1(n_1-1,0;k)d_2(n_1-1,0;k)\Bigr\}
  \nonumber\\
 &&+\sum_{n_1=1}^{\infty}\sum_{n_2=1}^{\infty}
  \cket{n_1}{n_2}\bra{n_1-1}{n_2+1}\nonumber\\
 &&\hspace{5mm}\times
  \Bigl\{d_1(n_1-1,n_2+1;k)d_2(n_1,n_2;k)
  -d_1(n_1-1,n_2;k)d_2(n_1-1,n_2;k)\Bigr\}\Biggr],\nonumber\\
\end{eqnarray}
where $d_1(n_1,n_2;k)$ and $d_2(n_1,n_2;k)$ are defined
by (\ref{EQ:d1}) and (\ref{EQ:d2}).
At a first sight, it is difficult to check whether the condition
$F_{1\bar{1}}=-F_{2\bar{2}}$ is satisfied or not.
However, we confirmed that the condition is satisfied really by using a
numerical evaluation.


\end{document}